\documentclass[aps,pra,reprint,superscriptaddress,longbibliography,floatfix]{revtex4-2}

\usepackage{amsmath,amssymb,bm,mathtools}
\usepackage{graphicx}
\usepackage{booktabs}
\usepackage{microtype}
\usepackage{placeins}
\usepackage{tikz}
\usetikzlibrary{arrows.meta,positioning}
\usepackage{xcolor}
\usepackage{hyperref}
\hypersetup{hidelinks}
\usepackage{extarrows}
\newcommand{\dd}{\mathrm{d}}
\newcommand{\ii}{\mathrm{i}}
\newcommand{\Tr}{\operatorname{Tr}}
\newcommand{\D}{\mathcal{D}}
\newcommand{\Lc}{\mathcal{L}}

\newcommand{\eps}{\varepsilon}
\newcommand{\pos}[1]{\left[#1\right]_{+}}

\begin{document}

\title{A Nonrecursive Lindblad Quantization of Dissipative Polynomial Dynamics}

\author{Tingfei Li}
\email{tfli@hbu.edu.cn}
\affiliation{College of Physics Science and Technology, Hebei University, Baoding 071002, China}
\affiliation{Hebei Key Laboratory of High-Precision Computation and Application of Quantum Field Theory, Baoding 071002, China}
\affiliation{Hebei Research Center of the Basic Discipline for Computational Physics, Baoding 071002, China}

\date{\today}

\begin{abstract}
	We present a direct and nonrecursive construction that maps an arbitrary planar polynomial dissipative flow
	$\dot{\alpha}=h(\alpha,\alpha^*)$
	to an open quantum system in Gorini--Kossakowski--Sudarshan--Lindblad form. Given the homogeneous components of the classical drift, the corresponding Hamiltonian and collapse-operator blocks are obtained algebraically and independently at each degree. No recursive cancellation of lower-order terms is required: in the large-amplitude limit $|\alpha|\sim S\to\infty$, ordering-generated lower-degree terms are parametrically suppressed, while mean-field closure yields the prescribed leading $O(S)$ drift for semiclassically localized states. This provides a simple and systematic route from dissipative classical dynamics to explicit open-quantum-system realizations, without claiming a unique microscopic quantization of the classical flow.
	We demonstrate the construction for stable fixed points, a Hopf bifurcation, and a bistable-ring flow. The corresponding Liouvillian spectra recover the classical relaxation exponents, the radial Floquet exponent and neutral phase direction of a limit cycle, and the separation between local relaxation and inter-ring switching in the bistable case. These examples show that the construction can be used not only to analyze a given quantum model, but also to design open quantum systems with prescribed classical dynamical structures in the semiclassical limit.
\end{abstract}


\maketitle

\section{Introduction}

Dissipation is not a small correction to idealized dynamics; it selects steady states, sets relaxation time scales, stabilizes self-sustained oscillations, and enables bifurcations, multistability, and nonequilibrium phase transitions \cite{Strogatz2000,CrossHohenberg1993,Haken1975}.  These structures are well organized in classical dynamical-systems theory, where vector fields and their invariant sets provide a compact language for a wide variety of physical, chemical, and biological phenomena.  Quantum dissipation is equally fundamental: laboratory quantum systems inevitably interact with their surroundings, and engineered reservoirs can create states and dynamical phases unavailable in closed systems \cite{BreuerPetruccione,GardinerZoller,Poyatos1996,Verstraete2009}.  Developing quantum counterparts of classical dissipative systems is therefore important both for understanding how familiar nonlinear dynamics emerges from quantum mechanics and for designing open quantum systems with prescribed macroscopic behavior.

A productive route is to reverse the usual direction of analysis.  Instead of starting from a microscopic quantum model and asking for its classical limit, one starts from a desired classical flow and constructs an open quantum dynamics that realizes it.  This inverse-design viewpoint underlies reservoir engineering and recent constructive approaches to non-Hamiltonian dynamics \cite{Poyatos1996,Verstraete2009,Chia2025}.  The classical system then serves as a design specification: its fixed points, cycles, bifurcations, and basins identify the structures that the quantum model should recover at macroscopic scales, while the quantum realization makes fluctuations, phase diffusion, switching, and Liouvillian relaxation accessible beyond a deterministic mean-field description.

The difficulty of quantizing dissipation has been recognized since the early development of quantum theory.  For the damped harmonic oscillator, time-dependent canonical Hamiltonians and enlarged conservative formulations provided influential constructions, but also showed that inequivalent Hamiltonian embeddings can reproduce the same damped equation \cite{Kanai1948,Dekker1981}.  System--reservoir models instead quantize a larger closed system and derive dissipation after eliminating environmental degrees of freedom; their predictions necessarily depend on the chosen coupling and bath spectral density \cite{CaldeiraLeggett1983,BreuerPetruccione}.  Other approaches have represented classical evolution in Hilbert space or proposed direct quantization rules for non-Hamiltonian flows \cite{Koopman1931,vonNeumann1932,Joseph2020,Tarasov2004,Vol2006}.  These efforts establish that dissipative dynamics can be embedded in quantum form in several ways, but they also expose the central ambiguity: the reduced classical vector field alone does not select one of these quantum completions.

Cascade quantization provides a general constructive implementation of this idea for polynomial planar flows by realizing successive monomials with Hamiltonian and dissipative channels \cite{Chia2025}.  Its exact version, however, recursively cancels lower-degree terms generated at each stage.  For high-degree vector fields this recursion makes the construction increasingly cumbersome and proliferates operator terms and dissipative channels.  More importantly, exact cancellation is tied to a chosen operator ordering.  It removes ordering-generated contributions to the first-moment equation while the mean-field replacement still neglects connected quantum moments of comparable subleading significance.  Exact agreement within one ordering prescription therefore does not, by itself, provide a controlled measure of finite-size accuracy.

This observation motivates the central viewpoint of this work: a classical dissipative vector field does not define a unique, microscopically ``correct'' quantum theory.  It specifies neither bath correlations and a noise kernel nor system--environment couplings or an operator ordering.  Different open quantum systems may consequently share the same classical drift while having different finite-size fluctuations and spectra.  In the absence of additional microscopic information, the physically invariant target is thus not exact operator-level agreement at finite quantum scale.  It is sufficient---and, in general, all that can be required---that the quantum dynamics reduce to the prescribed classical system under mean-field closure in a well-defined classical limit \cite{Wigner1932,Moyal1949,Groenewold1946,VanHove1951,CahillGlauber1969a,CahillGlauber1969b,Hillery1984}.

We implement this principle for a polynomial flow on $\mathbb{R}^{2}$.  Introducing
\begin{equation}
 \alpha=\frac{x+\ii y}{\sqrt{2}},
 \qquad
 \dot\alpha=h(\alpha,\alpha^*),
 \label{eq:intro-classical-flow}
\end{equation}
we seek a single-mode open quantum system with $[a,a^\dagger]=1$ and a Gorini--Kossakowski--Sudarshan--Lindblad (GKSL) generator $\mathcal{L}$. 
Markovian open-system dynamics supplies a natural algebraic setting because GKSL form preserves complete positivity and trace at every finite Fock-space cutoff \cite{Kraus1971,Kossakowski1972,Davies1974,Gorini1976,Lindblad1976,Evans1977}.  We identify the classical amplitude with the first moment, $\alpha=\langle a\rangle$, and introduce a macroscopic scale $|\alpha|\sim S\gg1$.  Equivalently, the scaled operator $b=a/S$ obeys $[b,b^\dagger]=S^{-2}$, so $S\to\infty$ is a small-effective-Planck-constant limit.

Our matching criterion keeps the order of operations explicit.  We first evaluate the adjoint evolution $\mathcal L^\dagger a$ exactly, then apply the mean-field replacement
\begin{equation}
 \langle(a^\dagger)^m a^n\rangle
 \longrightarrow(\alpha^*)^m\alpha^n,
\end{equation}
and finally retain the leading term under $|\alpha|\sim S\to\infty$.  The required correspondence is
\begin{equation}
 \left\langle\mathcal L^\dagger a\right\rangle_{\rm MF}
 =h(\alpha,\alpha^*)+o(S).
 \label{eq:intro-matching}
\end{equation}
Homogeneous operator degrees are parametrically separated in this limit, as in system-size expansions and Wigner--Moyal calculus \cite{vanKampen1961,KuboMatsuoKitahara1973,DrummondGardiner1980,Risken}.  Operator-ordering contractions enter below the leading layer algebraically, as follows from the normal-product formula in Appendix~\ref{app:algebra}.  Connected moments discarded by mean field are a separate, state-dependent source of corrections: they are subleading for suitably localized semiclassical states but need not be small for broad or multimodal states.  Keeping this distinction explicit avoids assigning artificial privilege to one finite-$S$ ordering prescription.

On this basis, we develop a direct, nonrecursive version of cascade quantization.  We decompose $h=\sum_n h_n$ into homogeneous polynomials and construct the Hamiltonian and jump-operator block for every $h_n$ algebraically and independently.  Each block reproduces its target component at leading order, while all induced lower-degree terms vanish relative to the $O(S)$ macroscopic drift.  No lower-order cancellation cascade is needed.  The resulting prescription is modular, transparent in its large-$S$ power counting, and applicable to an arbitrary polynomial planar flow.  It also makes the scope of the claim precise: the construction supplies a controlled representative of a semiclassical correspondence, rather than a unique microscopic quantization of the classical equation.

The polynomial Hamiltonians and jump operators should be understood as formal infinite-dimensional GKSL expressions unless appropriate domain and growth conditions are imposed \cite{ChebotarevFagnola1998,BahnKoPark2005,SiemonHolevoWerner2017}.  At finite Fock cutoff, which is the setting of our numerical calculations, the generator is manifestly completely positive and trace preserving.  This distinction separates the constructive algebraic result established here from the stronger problem of proving existence and conservativity of a quantum dynamical semigroup on the full bosonic Fock space.

We test more than the first-moment drift.  Phase-space distributions and Liouvillian spectra probe whether the constructed model recovers classical invariant sets together with their stability and relaxation scales.  Wigner methods connect the quantum steady state to the underlying phase-space geometry \cite{Glauber1963,Sudarshan1963,CahillGlauber1969b,Hillery1984}; rapidity methods and symmetry-resolved Liouville sectors organize the spectrum \cite{Prosen2008,Prosen2010,ProsenSeligman2010,BarthelZhang2022,McDonaldClerk2023,AlbertJiang2014}; and separated low-lying eigenvalues diagnose metastability and dissipative criticality \cite{Kessler2012,Macieszczak2016,Minganti2018,Macieszczak2021,Minganti2021}.

Open-system time crystals make this connection particularly concrete.  Theoretical studies have identified persistent oscillatory phases in boundary-driven many-body systems, symmetry-supported dissipative dynamics, and lossy Bose--Hubbard models \cite{Iemini2018,Buca2019,Lledo2019}, and dissipative time-crystalline order has been observed in an atom--cavity experiment \cite{Kessler2021TimeCrystal}.  In spectral language, such macroscopic oscillations require Liouvillian modes whose decay rates vanish in the appropriate thermodynamic or classical limit.  Closely related gap closing produces critical slowing down near dissipative phase transitions: it has been analyzed in nonlinear resonators and driven-dissipative Bose--Hubbard lattices \cite{Casteels2017,Vicentini2018}, observed in circuit quantum electrodynamics \cite{Brookes2021}, and connected directly to the quantum origin of classical fixed points, limit cycles, and Hopf criticality \cite{DuttaZhangHaque2025}.  These works demonstrate that limit cycles and critical slowing are central organizing phenomena in open quantum systems and motivate our use of the Hopf gap, radial Floquet modes, and the neutral phase branch as benchmarks of the constructed classical limit.

More broadly, self-sustained oscillations have a long classical history \cite{Rayleigh1883,VanDerPol1926,Winfree1967,Kuramoto1975,Strogatz2000,CrossHohenberg1993,Haken1975}.  Their quantum counterparts have also been studied through van der Pol and Rayleigh oscillators, synchronization, phase reduction, and fluctuation theories around limit cycles \cite{LeeSadeghpour2013,Walter2014,Walter2015,Lorch2014,NavarreteBenlloch2017,RouletBruder2018,KoppenhoferRoulet2019,Kato2019,Kato2020,BenArosh2021,Zhu2022,Setoyama2024,Sudler2024,delPino2024,NadolnyBruder2026}.  This literature provides complementary tools for distinguishing deterministic attraction toward a cycle from quantum phase diffusion along it.

\begin{figure*}[t]
\centering
\begin{tikzpicture}[
  node distance=8mm and 9mm,
  every node/.style={font=\normalsize},
  box/.style={draw, rounded corners, align=center, minimum width=29mm, minimum height=12mm, inner sep=5pt, line width=0.9pt},
  arrow/.style={-{Latex[length=2.8mm,width=2.0mm]}, line width=1.0pt},
  annotation/.style={font=\small, align=center}
]
\node[box] (flow) {Classical flow\\$\dot\alpha=\sum_{n=0}^{N}h_n$};
\node[box, right=of flow] (scale) {Homogeneous hierarchy\\$|\alpha|\sim S\gg1$};
\node[box, right=of scale] (gksl) {Order-$n$ blocks\\$H_n,\{J_{n,k}\}$};
\node[box, below=of scale] (wigner) {Wigner generator\\drift, diffusion, and\\higher derivatives};
\node[box, left=of wigner] (fixed) {Stable fixed point\\quadratic benchmark};
\node[box, right=of wigner] (cycle) {Limit cycles\\radial, phase, and\\switching modes};
\node[box, below=of wigner] (spectrum) {Liouvillian spectrum\\exact diagonalization\\versus asymptotics};

\draw[arrow] (flow) -- (scale);
\draw[arrow] (scale) -- (gksl);
\draw[arrow] (gksl) -- (wigner);
\draw[arrow] (wigner) -- (fixed);
\draw[arrow] (wigner) -- (cycle);
\draw[arrow] (fixed) -- (spectrum);
\draw[arrow] (cycle) -- (spectrum);
\node[annotation, above=1.5mm of gksl] {explicit leading-order construction};
\end{tikzpicture}
\caption{Construction and large-$S$ validation. A polynomial flow is decomposed into homogeneous components, and explicit formal GKSL blocks reproduce each component at leading order. Mean-field reduction gives the classical drift as $S\to\infty$. Wigner analysis and symmetry-resolved diagonalization then test classical fixed points, limit cycles, radial relaxation, and the suppression of inter-ring switching.}
\label{fig:scheme}
\end{figure*}
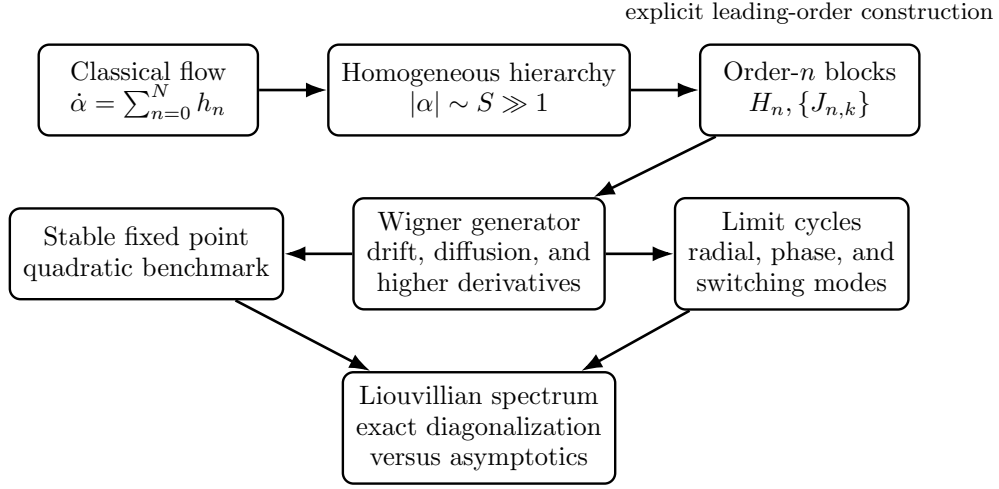

The paper is organized as follows.  Section~\ref{sec:construction} defines the large-$S$ correspondence and derives explicit even- and odd-degree blocks.  Section~\ref{sec:tests} tests the construction using a stable fixed point, a Hopf bifurcation and its limit-cycle phase, and a bistable pair of limit cycles.  The final Discussion and Outlook section summarizes the results, clarifies their asymptotic scope, and outlines future extensions.  Appendix~\ref{app:algebra} collects the normal-ordering algebra, Appendices~\ref{app:even} and \ref{app:odd} give the detailed even- and odd-degree block derivations, and Appendix~\ref{app:numerics} records the charge-sector formulas used in the numerical tests.  Figure~\ref{fig:scheme} summarizes this workflow; the accompanying Python and Mathematica code reproduces the numerical results.

\section{Explicit semiclassical cascade construction}
\label{sec:construction}

\subsection{Conventions and the meaning of the expansion}
We now make the scaling and matching prescription precise.  Let the drift in Eq.~\eqref{eq:intro-classical-flow} be a polynomial of maximal degree $N$, parameterized as
\begin{equation}
 h(\alpha,\alpha^*)=
 S\sum_{i+j=0}^{N}\frac{c_{ij}}{S^{i+j}}
 \alpha^{*i}\alpha^j,
 \qquad |\alpha|\sim S.
 \label{eq:scaled-polynomial}
\end{equation}
This scaling keeps every homogeneous component of the classical drift at order $S$ when the scaled amplitude $\alpha/S$ is fixed.  Throughout the paper we set $\hbar=1$; as noted in the Introduction, the effective semiclassical parameter is instead $S^{-2}$.  We promote $\alpha$ and $\alpha^*$ to a bosonic pair $a$ and $a^\dagger$ satisfying $[a,a^\dagger]=1$.  For an abstract vector field this is a construction of an open quantum system with analogous macroscopic dynamics, not a claim that $x$ and $y$ possess a unique canonical interpretation.

We use the standard GKSL equation
\begin{equation}
	\dot\rho=-\ii[H,\rho]+\sum_\nu \D[J_\nu]\rho,
	\qquad
	\D[J]\rho=J\rho J^\dagger-\frac12\{J^\dagger J,\rho\}.
	\label{eq:gksl}
\end{equation}
Its adjoint action is
\begin{align}
	\Lc^\dagger O&=\ii[H,O]+\sum_\nu\D_{J_\nu}^\dagger(O),\nonumber\\
	\D_J^\dagger(O)&=J^\dagger OJ-\frac12\{J^\dagger J,O\}.
	\label{eq:adjoint}
\end{align}
Equations~\eqref{eq:gksl} and \eqref{eq:adjoint} define a formal algebraic GKSL realization for polynomial $H$ and $J_\nu$. At every finite Fock cutoff the resulting matrix generator is manifestly completely positive and trace preserving. For unbounded polynomial operators on the full Fock space, existence and conservativity require a common invariant domain and additional growth conditions \cite{ChebotarevFagnola1998,BahnKoPark2005,SiemonHolevoWerner2017}; no universal infinite-dimensional semigroup theorem is asserted here.
  We write
\begin{equation}
 \langle A\rangle_t\equiv\Tr[A\rho(t)]
\end{equation}
and suppress the time subscript when no confusion can arise.

In this work, ``quantization'' means finding $H$ and a set of jump operators $\{J_\nu\}$ such that, after the exact adjoint action is evaluated, mean-field factorization is applied, and the leading large-$S$ term is retained,
\begin{equation}
 \frac{\dd}{\dd t}\langle a\rangle
 =\left\langle\ii[H,a]+\sum_\nu\D_{J_\nu}^\dagger(a)\right\rangle
 \xlongequal[\phantom{S\to\infty}]{\mathrm{MF},\,S\to\infty}
 h(\alpha,\alpha^*),
 \label{eq:op-dyn}
\end{equation}
where $\alpha=\langle a\rangle$ and mean-field factorization replaces normal-ordered moments by products of first moments.  The order of these operations is essential: the operator evolution is evaluated before factorization and large-$S$ power counting.  Below, the symbol $\asymp$ denotes equality under this combined mean-field and leading-order prescription.  This is an algebraic matching criterion; convergence of the exact moment dynamics additionally requires a family of states for which connected moments are subleading.

We decompose the polynomial as $h=\sum_{n=0}^N h_n$, where $h_n$ is homogeneous of degree $n$, and construct the corresponding Hamiltonian $H_n$ and jumps $\{J_{n,k}\}$ independently.  The normal-ordering identities used throughout this construction are collected in Appendix~\ref{app:algebra}:
\begin{equation}
 \left\langle\ii[H_n,a]+\sum_k\D_{J_{n,k}}^\dagger(a)\right\rangle
 \asymp h_n.
 \label{eq:block-matching}
\end{equation}

Write the degree-$n$ component as
\begin{align}
 h_n=\frac{1}{S^{n-1}}\Bigg[&\lambda_n(\alpha^*)^n
 +\sum_{k=0}^{K(n)}\mu_{n,k}(\alpha^*)^k\alpha^{n-k}
 \nonumber\\[-2pt]
 &+\sum_{k=0}^{K(n)}\nu_{n,k}
 (\alpha^*)^{n-k-1}\alpha^{k+1}\Bigg],
 \label{eq:hn}
\end{align}
where
\begin{align}
 K(2m)&=m-1,\qquad K(2m+1)=m.
 \label{eq:Kdefinition}
\end{align}
For odd degree $n=2m+1$, the two monomials at $k=m$ coincide; their total coefficient is
\begin{equation}
 C_m\equiv \mu_{2m+1,m}+\nu_{2m+1,m}.
 \label{eq:Cm}
\end{equation}
The complete drift is $h=\sum_{n=0}^{N}h_n$. 
The pure conjugate monomial $\lambda_n \alpha^{*n}/S^{n-1} $  is generated by
\begin{equation}
 H_n^{(\lambda)}=
 \frac{\ii}{(n+1)S^{n-1}}
 \left[\lambda_n(a^\dagger)^{n+1}-\lambda_n^*a^{n+1}\right],
 \label{eq:pure-block}
\end{equation}
which obeys $\ii[H_n^{(\lambda)},a]=S^{1-n}\lambda_n(a^\dagger)^n$ exactly.

\subsection{Even-degree blocks}

For $n=2m\geq2$ and $k=0,\ldots,m-1$, define
\begin{align}
 H_{2m,k}={}&\frac{1}{S^{2m-1}}
 \left[z_{2m,k}(a^\dagger)^{k+1}a^{2m-k}+\mathrm{H.c.}\right],
 \label{eq:even-H}\\
 J^{(1)}_{2m,k}={}&\sqrt{\frac{2}{S^{2m-1}}}
 \left[a^{m+1}+\xi_{2m,k}(a^\dagger)^{m-k-1}a^{k+1}\right],
 \label{eq:even-J1}\\
 J^{(g)}_{2m}={}&\sqrt{\frac{2m}{S^{2m-1}}}(a^\dagger)^{m+1}.
 \label{eq:even-Jg}
\end{align}
As derived in Appendix~\ref{app:even}, the leading mixed terms match the target coefficients provided that
\begin{subequations}
\begin{align}
 -(k+1)\xi_{2m,k}^*-\ii(k+1)z_{2m,k}&=\mu_{2m,k},\\
 -(k+2)\xi_{2m,k}-\ii(2m-k)z_{2m,k}^*&=\nu_{2m,k}.
\end{align}
\label{eq:even-match}
\end{subequations}
Solving Eq.~\eqref{eq:even-match} gives
\begin{subequations}
\begin{align}
 z_{2m,k}&=
 \frac{\ii[(k+2)\mu_{2m,k}-(k+1)\nu_{2m,k}^*]}
 {2(k+1)(m+1)},
 \label{eq:even-z}\\
 \xi_{2m,k}&=
 \frac{(k-2m)\mu_{2m,k}^*-(k+1)\nu_{2m,k}}
 {2(k+1)(m+1)}.
 \label{eq:even-xi}
\end{align}
\end{subequations}
The denominators are nonzero throughout the allowed range. The full even block is
\begin{align}
 H_{2m}&=H_{2m}^{(\lambda)}+\sum_{k=0}^{m-1}H_{2m,k},\nonumber\\
 \{J_{2m,\nu}\}_{\nu}&=
 \{J^{(1)}_{2m,k}\}_{k=0}^{m-1}\cup\{J^{(g)}_{2m}\}.
 \label{eq:even-full}
\end{align}
After the exact adjoint action is evaluated, all lower-degree terms are discarded and the mean-field identification gives
\begin{equation}
 \left\langle\Lc_{2m}^\dagger a\right\rangle\asymp h_{2m}.
 \label{eq:even-result}
\end{equation}
The cancellation of the leading self-actions and the degree counting of the omitted contractions are shown in Appendix~\ref{app:even}.

\subsection{Odd-degree blocks and the central completion}

For $n=2m+1\geq1$ and $k=0,\ldots,m-1$, define
\begin{align}
 H_{2m+1,k}={}&\frac{1}{S^{2m}}
 \left[z_{2m+1,k}(a^\dagger)^{k+1}a^{2m+1-k}+\mathrm{H.c.}\right],
 \label{eq:odd-H}\\
 J_{2m+1,k}={}&\sqrt{\frac{2}{S^{2m}}}
 \left[a^{m+1}+\xi_{2m+1,k}(a^\dagger)^{m-k}a^{k+1}\right].
 \label{eq:odd-J}
\end{align}
The mixed matching equations, derived explicitly in Appendix~\ref{app:odd}, are
\begin{subequations}
\begin{align}
 -(k+1)\xi_{2m+1,k}^*-\ii(k+1)z_{2m+1,k}&=\mu_{2m+1,k},\\
 -(k+1)\xi_{2m+1,k}-\ii(2m+1-k)z_{2m+1,k}^*&=\nu_{2m+1,k},
\end{align}
\label{eq:odd-match}
\end{subequations}
with solution
\begin{subequations}
\begin{align}
 z_{2m+1,k}&=
 \frac{\ii(\mu_{2m+1,k}-\nu_{2m+1,k}^*)}{2(m+1)},
 \label{eq:odd-z}\\
 \xi_{2m+1,k}&=
 \frac{(k-2m-1)\mu_{2m+1,k}^*-(k+1)\nu_{2m+1,k}}
 {2(k+1)(m+1)}.
 \label{eq:odd-xi}
\end{align}
\end{subequations}
The self-actions of Eq.~\eqref{eq:odd-J} contribute to the central monomial.  Appendix~\ref{app:odd} also derives the completion of this remaining coefficient.  Define
\begin{align}
 \beta_k&=m+1+|\xi_{2m+1,k}|^2(2k+1-m),\nonumber\\
 \eps_{2m+1}&=C_m+\sum_{k=0}^{m-1}\beta_k.
 \label{eq:epsilon}
\end{align}
The imaginary and real parts are completed by
\begin{align}
 H_{2m+1}^{(c)}={}&-
 \frac{\Im\eps_{2m+1}}{(m+1)S^{2m}}
 (a^\dagger)^{m+1}a^{m+1},
 \label{eq:central-H}\\
 J_{2m+1}^{(+)}={}&
 \sqrt{\frac{2\pos{\Re\eps_{2m+1}}}{(m+1)S^{2m}}}
 (a^\dagger)^{m+1},
 \label{eq:central-Jplus}\\
 J_{2m+1}^{(-)}={}&
 \sqrt{\frac{2\pos{-\Re\eps_{2m+1}}}{(m+1)S^{2m}}}
 a^{m+1}.
 \label{eq:central-Jminus}
\end{align}
where $[x]_+=\max\{0,x\}$, so at most one of Eqs.~\eqref{eq:central-Jplus} and \eqref{eq:central-Jminus} is nonzero. Nonnegative coefficients are manifest at finite cutoff, and the generator is continuous (not generally differentiable) at $\Re\eps_{2m+1}=0$. The full odd block is
\begin{equation}
 H_{2m+1}=H_{2m+1}^{(\lambda)}+
 \sum_{k=0}^{m-1}H_{2m+1,k}+H_{2m+1}^{(c)},
 \label{eq:odd-full-H}
\end{equation}
with the jumps in Eqs.~\eqref{eq:odd-J}, \eqref{eq:central-Jplus}, and \eqref{eq:central-Jminus}. Its leading drift equals $h_{2m+1}$; all remaining terms have lower degree and are dropped in the large-$S$ mean-field limit.  The self-action coefficients, central completion, and contraction power counting underlying this conclusion are given in Appendix~\ref{app:odd}.

\section{Liouvillian tests}
\label{sec:tests}

\subsection{Stable fixed point: exact quadratic benchmark}
\begin{figure}[ht!]
	\centering
	\begin{minipage}{0.70\columnwidth}
		\centering
		\includegraphics[width=\linewidth]{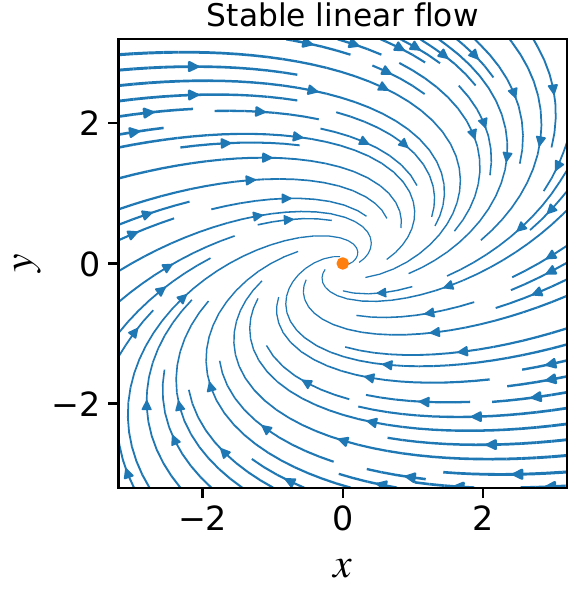}
	\end{minipage}\hfill
	\par\medskip
	\begin{minipage}{0.70\columnwidth}
		\centering
		\includegraphics[width=\linewidth]{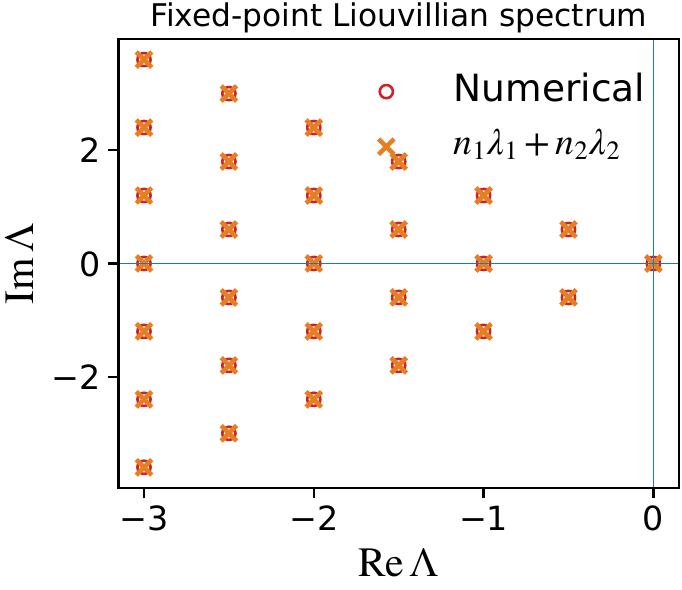}
	\end{minipage}
	\caption{Stable fixed-point benchmark. Top: classical linear flow. Bottom: low-lying Liouvillian eigenvalues from a Fock-space truncation at $N=30$ and the rapidity lattice generated by $-0.5\pm0.6\ii$.}
	\label{fig:fixed-point}
\end{figure}
Stable fixed points are the most elementary and ubiquitous local structures in dissipative dynamics: whenever a deterministic trajectory enters their basin, it approaches the fixed point exponentially.  They therefore provide a minimal test of whether the quantum construction retains the local relaxation data of the prescribed classical flow.  Linearizing about the fixed point gives
\begin{equation}
 \delta\bm r(t)=\sum_{j=1}^2 b_j e^{\lambda_j(C)t}\bm e_j .
 \label{eq:fp-classical-modes}
\end{equation}
Thus the eigenvalues of the classical stability matrix $C$ are the approach exponents: their real parts are decay rates, and a nonzero imaginary part describes a spiral into the fixed point.  We first treat the linear problem itself, where this connection can be made exactly rather than only semiclassically. Denote
\begin{equation}
 \frac{\dd}{\dd t}\begin{pmatrix}x\\y\end{pmatrix}
 =C\begin{pmatrix}x\\y\end{pmatrix},
 \qquad \max_j\Re\lambda_j(C)<0.
 \label{eq:linear-flow}
\end{equation}
In complex form,
\begin{equation}
 \dot\alpha=u\alpha+v\alpha^*,
 \label{eq:linear-complex}
\end{equation}
where
\begin{align}
 u&=\frac12\left[c_{11}+c_{22}+\ii(c_{21}-c_{12})\right],\\
 v&=\frac12\left[c_{11}-c_{22}+\ii(c_{12}+c_{21})\right].
\end{align}
The exact quadratic realization is
\begin{align}
 H_{\rm fp}&=\omega a^\dagger a+
 \frac{\ii}{2}\left[v(a^\dagger)^2-v^*a^2\right],
 \label{eq:fp-H}\\
 J_{\rm fp}&=\sqrt{\kappa}\,a,
 \qquad
 \kappa=-\Tr C,
 \qquad
 \omega=\frac{c_{12}-c_{21}}{2}.
 \label{eq:fp-J}
\end{align}
Direct substitution gives $\Lc_{\rm fp}^\dagger a=u a+v a^\dagger$ with no approximation. The Wigner equation is the probability-conserving Ornstein--Uhlenbeck equation \cite{Risken,GardinerZoller}
\begin{equation}
 \partial_tW=-\bm\nabla\!\cdot(C\bm r W)
 +\frac{\kappa}{4}\nabla^2W.
 \label{eq:fp-FPE}
\end{equation}
For a stable diagonalizable $C$ in the Gaussian spectral domain, the Liouvillian rapidities form
\begin{equation}
 \Lambda_{n_1n_2}=n_1\lambda_1+n_2\lambda_2,
 \qquad n_1,n_2\in\mathbb N_0.
 \label{eq:fp-spectrum}
\end{equation}
The first-order relaxation modes therefore carry precisely the two classical exponents $\lambda_1$ and $\lambda_2$.  Higher Liouvillian modes are their nonnegative integer sums: they describe relaxation of higher moments and Hermite--Gaussian distortions of the Wigner distribution.  Quantum noise fixes the stationary Gaussian width, but it does not shift these rapidities in this quadratic Ornstein--Uhlenbeck problem.  Hence the low-lying Liouvillian spectrum retains the classical local stability data---decay and, when present, rotation---around the stable point.  This exact correspondence is special to the quadratic benchmark; it is not a claim that a general nonlinear completion has a uniquely determined finite-$S$ spectrum.

This is a benchmark, not a principal novelty.  We take
\begin{equation}
 C=\begin{pmatrix}-0.7&1.0\\-0.4&-0.3\end{pmatrix},
 \qquad \lambda_\pm=-0.5\pm0.6\ii.
 \label{eq:fp-parameters}
\end{equation}
For the numerical comparison, we represent operators in the number-state basis $\{|0\rangle,\ldots,|N-1\rangle\}$ and diagonalize the resulting $N^2\times N^2$ Liouvillian matrix.  The spectrum in Fig.~\ref{fig:fixed-point} uses $N=30$; among the displayed modes with $\Re\Lambda>-3.1$, the largest distance to the rapidity lattice is $1.3\times10^{-8}$, and the stored trace-preservation error is zero.  Thus, ``cutoff'' means retaining only the first $N$ Fock states; it is a numerical truncation of the infinite-dimensional oscillator Hilbert space, whose boundary can slightly displace finite-cutoff eigenvalues.  Increasing $N$ restores the low-lying rapidities in Eq.~\eqref{eq:fp-spectrum}.  Because the adjoint drift is exact in this quadratic example, the first-order quantum relaxation exponents coincide with the classical stability exponents without invoking mean-field factorization or a large-$S$ limit.  The vectorization and symmetry-resolved matrix conventions used in this and the subsequent diagonalizations are specified in Appendix~\ref{app:numerics}.

\subsection{Hopf bifurcation: fixed point, critical point, and cycle}
\label{sec:hopf}
\begin{figure*}[t]
	\centering
	\includegraphics[width=0.98\textwidth]{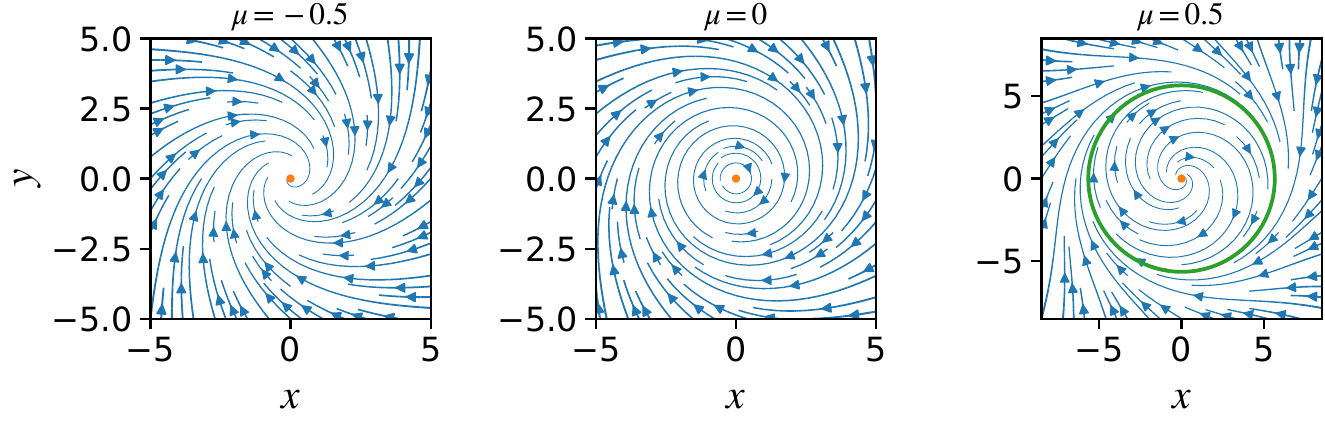}
	\includegraphics[width=0.98\textwidth]{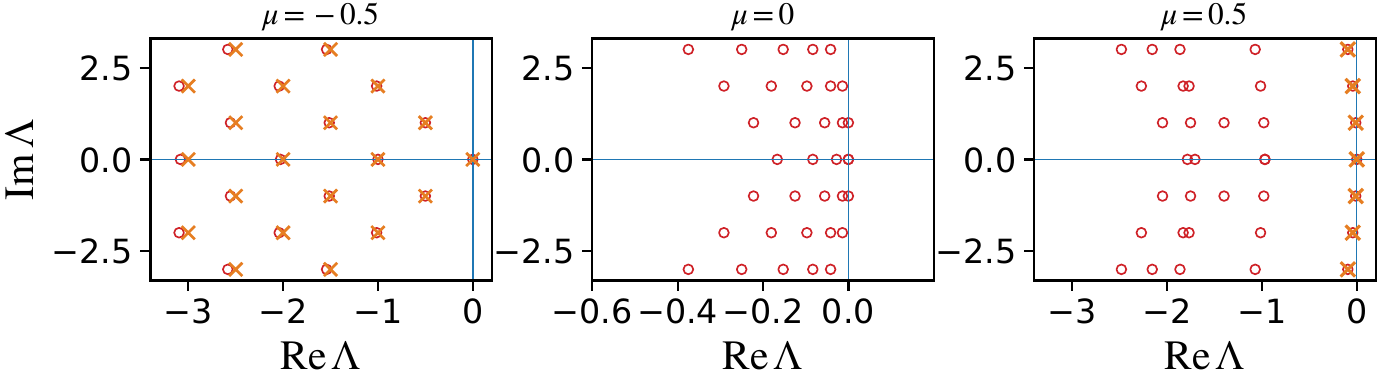}
\caption{Hopf scan for the minimal cascade completion.  Top: classical portraits below, at, and above the bifurcation.  Bottom: finite-$S$ spectra at $S=12$. Red open circles are exact charge-resolved quantum-Liouvillian eigenvalues; yellow crosses give the large-$S$ asymptotic values away from the critical point.}
	\label{fig:hopf-regimes}
\end{figure*}

A Hopf bifurcation is the canonical route by which a stable equilibrium loses stability and a self-sustained oscillation is born.  When a complex-conjugate pair of stability exponents crosses the imaginary axis, the deterministic dynamics changes from relaxation toward a fixed point to motion on a periodic orbit.  In the supercritical case considered here, this orbit is stable and its radius grows continuously from zero above threshold.  Hopf's original local bifurcation theory and its later role in nonlinear dynamics make this structure a standard organizing principle for oscillatory phenomena in physics, chemistry, biology, and engineering \cite{Hopf1942,Strogatz2000,Winfree1967,Kuramoto1975}.

For the minimal construction, the three regimes illustrate distinct aspects of the correspondence. Below threshold, the linear loss channel gives a finite relaxation gap associated with the stable origin. Above threshold, the gain channel together with two-photon loss produces a slow phase branch: radial perturbations relax toward the cycle, whereas the phase is neutral in the deterministic classical dynamics and acquires finite-$S$ phase diffusion. At the critical point the minimal construction has no linear channel, so it does not supply the independent linear noise needed for diffusive critical scaling. Consider the dynamical system below
\begin{equation}
	\dot\alpha=(\mu-\ii\omega)\alpha
	-\frac{2}{S^2}|\alpha|^2\alpha,
	\label{eq:hopf-alpha}
\end{equation}
or, equivalently,
\begin{equation}
	\dot r=r\left(\mu-\frac{r^2}{S^2}\right),
	\qquad \dot\phi=-\omega,
	\label{eq:hopf-polar}
\end{equation}
where $r^2=x^2+y^2=2|\alpha|^2$.  Hence the origin is stable for $\mu<0$, critical at $\mu=0$, and unstable for $\mu>0$, where a stable ring occurs at $r_*=S\sqrt\mu$.
We now apply the construction of Sec.~\ref{sec:construction} \emph{to this
classical equation}.  The degree-one component is
\begin{equation}
 h_1=(\mu-\ii\omega)\alpha.
\end{equation}
It is the $m=0$ central odd block, with $C_0=\mu-\ii\omega$.  Using the central-completion construction derived in Appendix~\ref{app:odd}, Eqs.
\eqref{eq:central-H}--\eqref{eq:central-Jminus} therefore give
\begin{align}
 H_1&=\omega a^\dagger a,\nonumber\\
 J_{1,+}&=\sqrt{2[\mu]_+}\,a^\dagger,
 \qquad J_{1,-}=\sqrt{2[-\mu]_+}\,a.
 \label{eq:hopf-minimal-linear}
\end{align}
The degree-three component is
\begin{equation}
 h_3=-\frac{2}{S^2}\alpha^*\alpha^2.
\end{equation}
It is the $m=1$ central odd block, with $C_1=-2$ and hence
$\eps_3=-2$.  Thus $H_3=0$, $J_{3,+}=0$, and
\begin{equation}
 J_{3,-}=\frac{\sqrt2}{S}a^2.
 \label{eq:hopf-minimal-nonlinear}
\end{equation}
The Hamiltonian and the nonzero operators in
Eqs.~\eqref{eq:hopf-minimal-linear}--\eqref{eq:hopf-minimal-nonlinear}
are used in every numerical calculation below. Their adjoint action is
\begin{equation}
 \Lc^\dagger a=(\mu-\ii\omega)a-\frac{2}{S^2}a^\dagger a^2.
 \label{eq:hopf-exact-drift}
\end{equation}
Thus the mean-field identification $\alpha=\langle a\rangle$ gives the prescribed leading macroscopic drift in Eq.~\eqref{eq:hopf-alpha}.

For completeness, we first state the Wigner-function convention used below.  For
\begin{equation}
 D(\lambda)=\exp(\lambda a^\dagger-\lambda^*a),
 \qquad
 \chi_W(\lambda,t)=\Tr\!\left[\rho(t)D(\lambda)\right],
\end{equation}
we define the Weyl-ordered Wigner function by
\begin{equation}
 W(\alpha,\alpha^*,t)
 =\frac{1}{\pi^2}\int\dd^2\lambda\,
 \exp\!\left(\alpha\lambda^*-\alpha^*\lambda\right)
 \chi_W(\lambda,t).
 \label{eq:wigner-definition}
\end{equation}
It follows directly that
$\int\dd^2\alpha\,W(\alpha,\alpha^*,t)=1$. Here $\alpha=(x+\ii y)/\sqrt{2}$, so that $\dd^2\alpha=\dd(\Re\alpha)\dd(\Im\alpha)$ and
$r^2=x^2+y^2=2|\alpha|^2$. With this convention, the ordered Bopp rules are
\begin{align}
 a\rho&\longleftrightarrow
 \left(\alpha+\frac12\partial_{\alpha^*}\right)W,
 &a^\dagger\rho&\longleftrightarrow
 \left(\alpha^*-\frac12\partial_\alpha\right)W,
 \nonumber\\
 \rho a&\longleftrightarrow
 \left(\alpha-\frac12\partial_{\alpha^*}\right)W,
 &\rho a^\dagger&\longleftrightarrow
 \left(\alpha^*+\frac12\partial_\alpha\right)W.
 \label{eq:hopf-bopp}
\end{align}
Keeping their order in every product gives the exact equation
\begin{align}
 \partial_tW={}&-\partial_\alpha(AW)
 -\partial_{\alpha^*}(A^*W)
 +\partial_\alpha\partial_{\alpha^*}(BW)
 \nonumber\\
 &+\frac{1}{2S^2}\left[
 \partial_\alpha^2\partial_{\alpha^*}(\alpha W)
 +\partial_\alpha\partial_{\alpha^*}^2(\alpha^*W)
 \right],
 \label{eq:hopf-wigner}
\end{align}
where
\begin{subequations}
\begin{align}
 A&=(\mu-\ii\omega)\alpha
 -\frac{2}{S^2}(|\alpha|^2-1)\alpha,
 \label{eq:hopf-wigner-A}\\
 B&=|\mu|+\frac{4|\alpha|^2-2}{S^2}.
 \label{eq:hopf-wigner-B}
\end{align}
\label{eq:hopf-wigner-coefficients}
\end{subequations}
In particular, the single linear channel gives
$A_{\rm lin}=\mu\alpha$ and $B_{\rm lin}=|\mu|$;
the two-photon channel supplies the remaining terms in
Eqs.~\eqref{eq:hopf-wigner-A} and \eqref{eq:hopf-wigner-B}, together with the last line of Eq.~\eqref{eq:hopf-wigner}.  Those third derivatives are genuine quantum Kramers--Moyal terms.  Consequently, Eq.~\eqref{eq:hopf-wigner} is not a second-order Fokker--Planck equation at finite $S$.

The macroscopic scaling makes the classical limit explicit.  With $\alpha=S\xi$ and $\int\dd^2\alpha\,W=1$, define the normalized distribution
\begin{equation}
 \widetilde W_S(\xi,\xi^*,t)
 =S^2W(S\xi,S\xi^*,t),
 \qquad \int\dd^2\xi\,\widetilde W_S=1.
 \label{eq:hopf-scaled-wigner}
\end{equation}
Equation~\eqref{eq:hopf-wigner} then becomes exactly
\begin{align}
 \partial_t\widetilde W_S={}&
 -\partial_\xi\!\left[
 \left(F+\frac{2\xi}{S^2}\right)\widetilde W_S\right]
 -\partial_{\xi^*}\!\left[
 \left(F^*+\frac{2\xi^*}{S^2}\right)\widetilde W_S\right]
 \nonumber\\
 &+\frac{1}{S^2}\partial_\xi\partial_{\xi^*}
 \left[\left(|\mu|+4|\xi|^2-\frac{2}{S^2}\right)
 \widetilde W_S\right]
 \nonumber\\
 &+\frac{1}{2S^4}
 \partial_\xi^2\partial_{\xi^*}(\xi\widetilde W_S)
 +\frac{1}{2S^4}
 \partial_\xi\partial_{\xi^*}^2(\xi^*\widetilde W_S),
 \label{eq:hopf-scaled-generator}\\
 F(\xi,\xi^*)&={}(\mu-\ii\omega)\xi
 -2|\xi|^2\xi.
 \label{eq:hopf-scaled-drift}
\end{align}
For fixed scaled coordinates and derivatives, the leading generator is therefore the deterministic Liouville generator
$-\partial_\xi(F\,\cdot)-\partial_{\xi^*}(F^*\,\cdot)$.
The diffusion and third-derivative terms are suppressed respectively by $S^{-2}$ and $S^{-4}$ in this scaling.  They characterize the chosen finite-$S$ completion; they are neither fixed by the classical drift nor interpreted as errors in its quantization.

The resulting low-lying spectrum has three controlled large-$S$ regimes.  For $\mu<0$, fluctuations remain $O(1)$ about the stable origin, so the nonlinear channel is suppressed and the limiting problem is a linear Ornstein--Uhlenbeck oscillator.  Its two conjugate rapidities generate
\begin{equation}
 \Lambda_{n_1n_2}
 =n_1(\mu-\ii\omega)+n_2(\mu+\ii\omega),
 \qquad n_1,n_2\in\mathbb N_0.
 \label{eq:hopf-origin-spectrum}
\end{equation}
The first-order modes are the two rotating decay directions of the linearized classical flow; their nonnegative integer sums describe higher polynomial or Hermite--Gaussian distortions of the stationary Wigner profile.  Thus the origin remains spectrally gapped below Hopf.

In the charge notation $l=n_2-n_1$, with $n=\min(n_1,n_2)$, the Wigner
Ornstein--Uhlenbeck eigenvalue equation gives
\begin{equation}
 \Lambda_{nl}^{\rm W,\,origin}
 =\mu(2n+|l|)+\ii l\omega,
 \qquad n\in\mathbb N_0,\quad l\in\mathbb Z.
 \label{eq:hopf-origin-nl}
\end{equation}
The relation between this angular label and the number-basis charge blocks used for exact diagonalization is stated in Appendix~\ref{app:numerics}.
These Wigner eigenvalues are used as the yellow crosses in the $\mu=-0.5$
panel of Fig.~\ref{fig:hopf-regimes}.

For $\mu>0$, set $r=r_*+\delta r$.  Linearizing Eq.~\eqref{eq:hopf-polar} gives
$\dot{\delta r}=-2\mu\,\delta r+O(S^{-1}\delta r^2)$.
Moreover, Eq.~\eqref{eq:hopf-wigner-B} gives
$B(r_*)=3\mu+O(S^{-2})$.  The leading backward radial--phase generator for $\delta r=O(1)$ is consequently
\begin{equation}
 \mathcal G_{\rm cyc}
 =\frac{3\mu}{2}\partial_{\delta r}^2
 -2\mu\,\delta r\,\partial_{\delta r}
 -\omega\partial_\phi+o(1).
 \label{eq:hopf-cycle-generator}
\end{equation}
The radial Ornstein--Uhlenbeck levels and phase harmonics therefore obey
\begin{equation}
 \Lambda_{nl}\longrightarrow-2\mu n+\ii  l\omega,
 \qquad n\in\mathbb N_0,\quad  l\in\mathbb Z.
 \label{eq:hopf-cycle-spectrum}
\end{equation}
The $n=1$ radial rate $-2\mu$ is precisely the transverse classical Floquet exponent.  Along the cycle the classical flow has no restoring force in phase, so the $n=0$ phase branch has vanishing real part as $S\to\infty$.

The leading finite-$S$ correction to the phase branch follows directly from the
fixed-charge quantum Liouvillian constructed with the matrix elements in Appendix~\ref{app:numerics}:
\begin{equation}
 \Lambda_{nl}^{\rm asym}
 =\ii l\omega-2\mu n+\frac{c_{nl}}{S^2}
 +O(S^{-4}),
 \qquad l\in\mathbb Z,\quad \mu>0.
 \label{eq:hopf-phase-correction}
\end{equation}
For the $n=0$ branch, the real-part coefficient in
$\Re\Lambda_{nl}=c_{nl}/S^2+O(S^{-4})$ is
\begin{equation}
 c_{0l}=-\frac{3}{2}l^2.
 \label{eq:hopf-c0l}
\end{equation}
This expression is compared with the exact fixed-charge quantum spectrum in the
right panel of Fig.~\ref{fig:hopf-regimes}: red open circles are the numerical
eigenvalues and yellow crosses are Eq.~\eqref{eq:hopf-phase-correction}.  For
$\mu=0.5$ and $S=12$, the largest absolute discrepancy over
$|l|\leq3$ is $3.44\times10^{-5}$, while it is $2.31\times10^{-6}$ for
$|l|=1$.  The remaining coefficients $c_{nl}$ with $n\geq1$ contain radial
quantum corrections and require the corresponding higher radial matrix elements
of the same fixed-charge quantum Liouvillian.

At $\mu=0$, the minimal construction contains only the two-photon loss channel.
Its even- and odd-parity dark populations give more than one stationary state,
so the model has neither a unique noisy critical steady state nor a conventional
primitive Liouvillian gap.  The middle panel of Fig.~\ref{fig:hopf-regimes} is
therefore retained only as a finite-cutoff spectrum and has no diffusive Wigner
overlay.

Figure~\ref{fig:hopf-regimes} compares the three regimes at
$\mu=-0.5,0,0.5$. The Wigner asymptotic values are obtained from
Eq.~\eqref{eq:hopf-origin-nl} below threshold and
Eq.~\eqref{eq:hopf-phase-correction} above threshold. These are leading
large-$S$, low-lying spectral statements, not an analytic diagonalization of
the full nonlinear finite-$S$ Liouvillian.  The charge-block calculations use
$N=163$ for $\mu=-0.5,0$ and $N=189$ for $\mu=0.5$; the largest stored
relative eigenpair residual is $1.57\times10^{-13}$.

\subsection{Bistable limit cycles: local relaxation and switching}
\begin{figure}[t]
	\centering
	\includegraphics[width=0.8\linewidth]{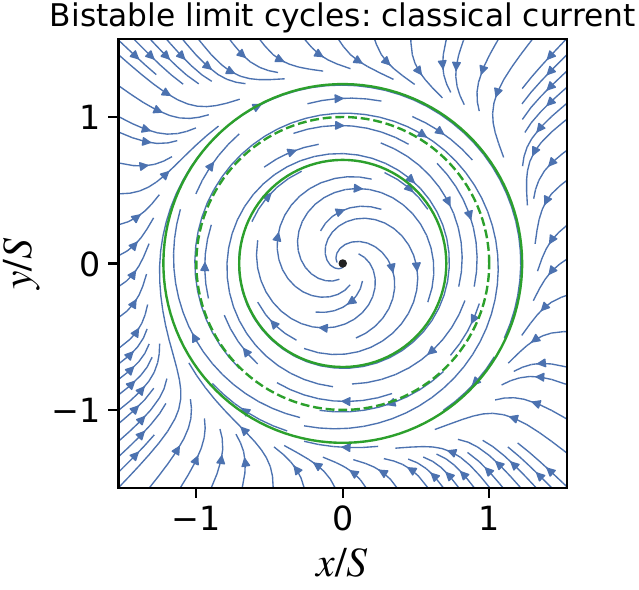}
	\caption{Classical current for the bistable limit-cycle flow with $(\mu_1,\mu_2,\mu_3)=(0.5,1,1.5)$ and $\omega=1$.  Arrows show the local velocity direction.  The green solid curves are stable limit cycles, whereas the green dashed curve is the unstable separatrix.  Trajectories inside and outside the unstable middle ring are driven toward the inner and outer stable limit cycles, respectively.}
	\label{fig:triple-current}
\end{figure}
\begin{figure*}[t]
	\centering
\begin{minipage}{0.33\textwidth}
	\centering
	\includegraphics[width=\linewidth]{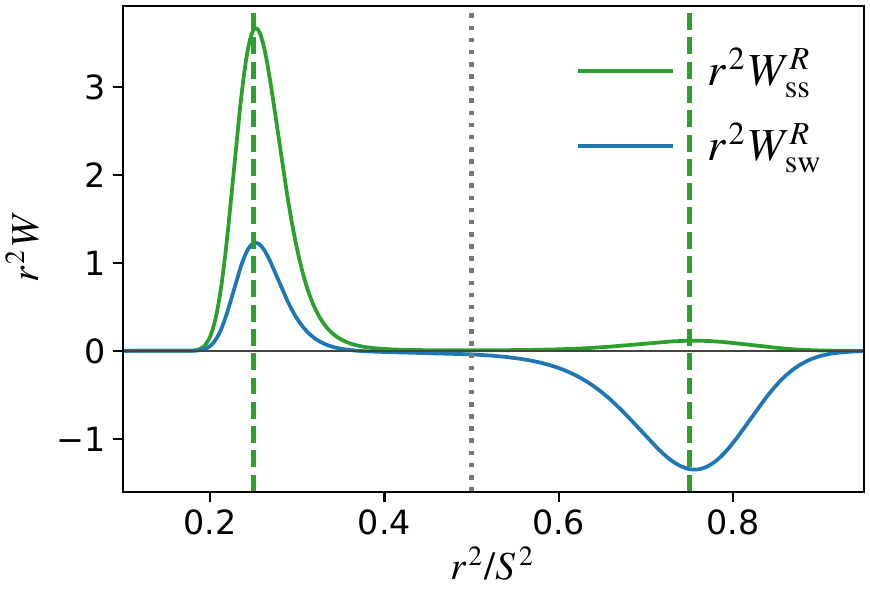}
\end{minipage}\hfill
\begin{minipage}{0.33\textwidth}
\centering
\includegraphics[width=\linewidth]{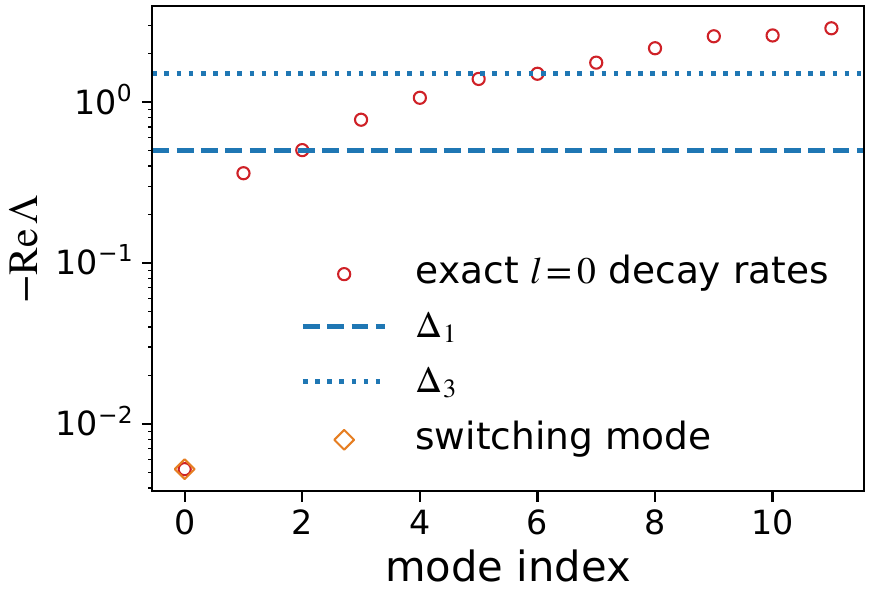}
\end{minipage}\hfill
\begin{minipage}{0.33\textwidth}
	\centering
	\includegraphics[width=\linewidth]{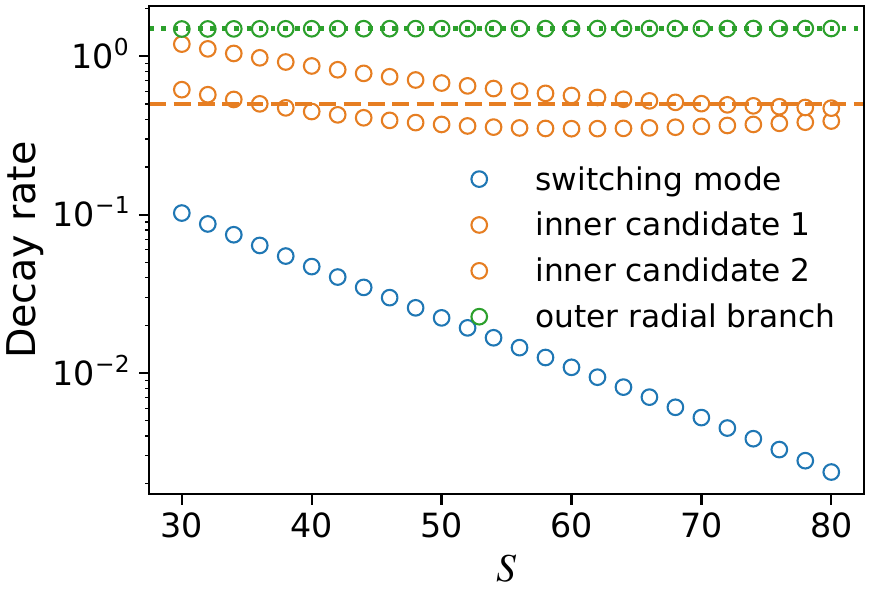}
\end{minipage}\hfill
	\caption{Bistable-ring structure at $S=70$. Left: the steady-state and switching right eigenfunctions are shown as $r^2W$ versus the common coordinate $r^2/S^2$, with $r=|\alpha|$; vertical lines mark the inner stable ring, the separatrix, and the outer stable ring. Middle: population-sector decay rates in the $l=0$ block, with the steady state at $\Lambda=0$ omitted. Hollow circles show the exact decay rates, the dashed and dotted horizontal lines mark the local gaps $\Delta_1$ and $\Delta_3$, and the hollow diamond marks the slower inter-ring switching mode. Right: scaling of the switching mode with $S$.}
	\label{fig:triple-main}
\end{figure*}
An important practical advantage of this construction is that it can be used to engineer dynamical phenomena directly.  Rather than guessing an open quantum system and then asking whether it supports bistability, one can prescribe a classical flow with two stable limit cycles and construct a corresponding Lindblad realization.  The method thus provides a route to open quantum systems with selected dynamical structures, rather than treating those structures as accidental features of an otherwise chosen model.

Consider
\begin{align}
 \dot r={}&-\frac{r}{S^6}
 (r^2-\mu_1S^2)(r^2-\mu_2S^2)(r^2-\mu_3S^2),
 \label{eq:triple-radial}
\end{align}
where $0<\mu_1<\mu_2<\mu_3$. 
The $S^{-6}$ factor in Eq.~\eqref{eq:triple-radial} is precisely what makes this scaled dynamics independent of $S$. Writing $s=r/S$ removes the macroscopic scale:
\begin{equation}
	\dot s=A(s),
	\qquad
	A(s)=-s\prod_{j=1}^3(s^2-\mu_j).
	\label{eq:triple-scaled-flow}
\end{equation}
The rings $s_1=\sqrt{\mu_1}$ and $s_3=\sqrt{\mu_3}$ are stable, while $s_2=\sqrt{\mu_2}$ is an unstable separatrix.  Classically, an initial condition inside or outside $s_2$ remains in that basin and approaches the corresponding stable cycle.  Figure~\ref{fig:triple-current} displays this radial stability structure together with the rotational current in the full plane.

With $e_3=\mu_1\mu_2\mu_3$, $e_2=\sum_{i<j}\mu_i\mu_j$, and $e_1=\sum_i\mu_i$, the equivalent complex flow is
\begin{align}
 \dot\alpha={}&(e_3-\ii\omega)\alpha
 -\frac{2e_2}{S^2}\alpha^*\alpha^2
 +\frac{4e_1}{S^4}(\alpha^*)^2\alpha^3
 -\frac{8}{S^6}(\alpha^*)^3\alpha^4.
 \label{eq:triple-alpha}
\end{align}
The GKSL realization used numerically is
\begin{align}
 H&=\omega a^\dagger a,
 \label{eq:triple-H}\\
 J_1&=\sqrt{2e_3}\,a^\dagger,
 &J_2&=\sqrt{\frac{2e_2}{S^2}}\,a^2,
 \nonumber\\
 J_3&=\sqrt{\frac{8e_1}{3S^4}}\,(a^\dagger)^3,
 &J_4&=\sqrt{\frac{4}{S^6}}\,a^4.
 \label{eq:triple-J}
\end{align}
The alternating gain and loss channels supply the successive signs of the radial polynomial: one-quantum gain destabilizes the origin, while nonlinear two- and four-quantum loss and three-quantum gain create the two saturation radii and the intervening separatrix.

After evaluating the adjoint action, applying the mean-field replacement, and dropping all terms suppressed for $|\alpha|\sim S\to\infty$, these operators give
\begin{equation}
 \frac{\dd}{\dd t}\langle a\rangle\asymp h(\alpha,\alpha^*),
\end{equation}
with $h$ equal to the classical drift in Eq.~\eqref{eq:triple-alpha}. No subleading matching is imposed.
Because all four drift monomials are central odd-degree monomials, this reduction is a direct application of the odd-block derivation in Appendix~\ref{app:odd}.

\paragraph{Exact finite-$S$ Wigner generator.---}
The finite-$S$ Wigner equation can be written exactly without displaying a
long seventh-order differential polynomial.  In addition to the Bopp
operators in Eq.~\eqref{eq:hopf-bopp}, define
\begin{align}
 a_L&=\alpha+\tfrac12\partial_{\alpha^*},
 &a_L^\dagger&=\alpha^*-\tfrac12\partial_\alpha,
 \nonumber\\
 a_R&=\alpha-\tfrac12\partial_{\alpha^*},
 &a_R^\dagger&=\alpha^*+\tfrac12\partial_\alpha .
 \label{eq:triple-bopp}
\end{align}
For a normally ordered monomial $J=c(a^\dagger)^p a^q$, let
\begin{align}
 J_L&=c(a_L^\dagger)^p a_L^q,
 &J_R&=c\,a_R^q(a_R^\dagger)^p,
 \nonumber\\
 (J^\dagger)_L&=c^*(a_L^\dagger)^q a_L^p,
 &(J^\dagger)_R&=c^*a_R^p(a_R^\dagger)^q .
\end{align}
The reversal in $J_R$ is essential because right multiplication is an
antirepresentation.  The exact Weyl image of a dissipator is then
\begin{equation}
 \mathcal D_W[J]
 =J_L(J^\dagger)_R-\frac12(J^\dagger)_LJ_L
 -\frac12J_R(J^\dagger)_R,
 \label{eq:triple-exact-dissipator}
\end{equation}
and the four-channel generator is exactly
\begin{equation}
 \mathcal L_W^{\rm exact}=-\ii\omega
 \left(a_L^\dagger a_L-a_Ra_R^\dagger\right)
 +\sum_{\nu=1}^4\mathcal D_W[J_\nu].
 \label{eq:triple-exact-wigner}
\end{equation}
Equations~\eqref{eq:triple-bopp}--\eqref{eq:triple-exact-wigner}, with the
$J_\nu$ in Eq.~\eqref{eq:triple-J}, give an exact finite-$S$ representation.
Upon expansion, derivatives through seventh order occur.  Thus the
exact equation is not a Fokker--Planck equation and no closed-form
diagonalization of the full nonlinear operator is known.

To expose the controlled macroscopic part, put $\alpha=S\xi$, define
$\widetilde W_S=S^2W(S\xi,S\xi^*)$, and write
$s=\sqrt2|\xi|$.  For distributions smooth on an $O(1)$ scale in $s$, the
forward equation obtained from Eq.~\eqref{eq:triple-exact-wigner} is
\begin{align}
 \partial_t\widetilde W_S={}&
 -\frac1s\partial_s\!\left[sA(s)\widetilde W_S\right]
 +\omega\partial_\phi\widetilde W_S
 +\frac{\partial_s\!\left[s\partial_s(B\widetilde W_S)\right]}
 {2S^2s}
 \nonumber\\
 &+\frac{\partial_\phi^2(B\widetilde W_S)}{2S^2s^2}
 +\frac1{S^2}\mathcal L_{\rm ord}\widetilde W_S
 +O_{\rm KM}(S^{-4}),
 \label{eq:triple-scaled-forward}\\
 B(s)={}&e_3+2e_2s^2+3e_1s^4+4s^6.
 \label{eq:triple-radial-noise}
\end{align}
Here $\mathcal L_{\rm ord}$ collects the first-derivative Weyl-ordering
corrections fixed by Eq.~\eqref{eq:triple-exact-wigner}, and
$O_{\rm KM}(S^{-4})$ starts with third derivatives.  These terms are
subleading for a smooth scaled Wigner function.  The leading backward
generator is therefore
\begin{equation}
 \mathcal G_s=A(s)\partial_s-\omega\partial_\phi
 +\frac{B(s)}{2S^2}
 \left(\partial_s^2+\frac1s\partial_s+
 \frac1{s^2}\partial_\phi^2\right)+\cdots .
 \label{eq:triple-scaled-generator}
\end{equation}
Its $S\to\infty$ limit is deterministic Liouville transport.  The diffusion
shown explicitly is nevertheless singularly important in $O(S^{-1})$
neighborhoods of a stable ring and for communication between basins.

The classical local radial gaps are
\begin{subequations}
\begin{align}
 \Delta_1&=2\mu_1(\mu_2-\mu_1)(\mu_3-\mu_1),\\
 \Delta_3&=2\mu_3(\mu_3-\mu_1)(\mu_3-\mu_2).
\end{align}
\label{eq:triple-gaps}
\end{subequations}
Let $x_i=r-Ss_i=S(s-s_i)$ near either stable ring, $i=1,3$, and
set $B_i=B(s_i)$.  Taylor expansion of the backward generator gives the
systematic boundary-layer hierarchy
\begin{equation}
 \mathcal G_i=\mathcal G_i^{(0)}+S^{-1}\mathcal G_i^{(1)}
 +S^{-2}\mathcal G_i^{(2)}+\cdots,
 \label{eq:triple-local-hierarchy}
\end{equation}
where
\begin{equation}
 \mathcal G_i^{(0)}=\frac{B_i}{2}\partial_{x_i}^2
 -\Delta_i x_i\partial_{x_i}-\omega\partial_\phi .
 \label{eq:triple-local-generator}
\end{equation}
For example, the diffusion-truncated part of the first correction is
\begin{equation}
 \left.\mathcal G_i^{(1)}\right|_{\rm diff}
 =\frac{A_i''}{2}x_i^2\partial_{x_i}
 +\frac{B_i'}{2}x_i\partial_{x_i}^2
 +\frac{B_i}{2s_i}\partial_{x_i},
 \label{eq:triple-local-first-correction}
\end{equation}
where primes are evaluated at $s_i$.  Weyl-ordering and the locally enhanced
third Kramers--Moyal derivative, both obtained by expanding
Eq.~\eqref{eq:triple-exact-wigner}, also enter $\mathcal G_i^{(1)}$.
The second correction contains, in particular,
$[B_i/(2s_i^2)]\partial_\phi^2$.  All terms in
$\mathcal G_i^{(1)}$ reverse radial parity.  Consequently their diagonal
matrix elements in the leading Ornstein--Uhlenbeck basis vanish and an
isolated fixed-$(n,l)$ eigenvalue has no $S^{-1}$ shift.

Define
\begin{equation}
 \sigma_i^2=\frac{B_i}{2\Delta_i},\qquad z_i=\frac{x_i}{\sigma_i}.
 \label{eq:triple-local-width}
\end{equation}
The leading right (forward) Wigner eigenfunctions and adjoint radial
polynomials are, up to normalization,
\begin{subequations}
\begin{align}
 W_{nl}^{(i),R}(r,\phi)
 &\simeq e^{+\ii l\phi}\,
 \operatorname{He}_n(z_i)e^{-z_i^2/2},
 \label{eq:triple-local-right}\\
 W_{nl}^{(i),L}(r,\phi)
 &\simeq e^{+\ii l\phi}\operatorname{He}_n(z_i).
 \label{eq:triple-local-left}
\end{align}
\label{eq:triple-local-functions}
\end{subequations}
The angular factor of the left mode is conjugated when the
Hilbert--Schmidt pairing is written explicitly.  Here
$\operatorname{He}_n$ denotes a probabilists' Hermite polynomial.  Thus the
fixed-$n$ modes really are local Ornstein--Uhlenbeck/Hermite modes; this
statement applies only while their radial support remains $o(S)$ away from
the separatrix.

The corresponding local spectrum is
\begin{align}
 \Lambda_{nl}^{(i)}={}&-n\Delta_i+\ii l\omega
 +\frac{c_{nl}^{(i)}}{S^2}+o(S^{-2}),
 \label{eq:triple-local-spectrum}\\
 c_{nl}^{(i)}={}&c_{n0}^{(i)}-\frac{B_i}{2\mu_i}l^2 .
 \label{eq:triple-local-correction}
\end{align}
Here $n\in\mathbb N_0$ and $l\in\mathbb Z$, with the convention $q=-l$.
The displayed $l^2$ term is the direct angular-diffusion contribution.  The
completion-dependent radial coefficient $c_{n0}^{(i)}$ contains a
second-order diagonal matrix element and two insertions of
$\mathcal G_i^{(1)}$ through other Hermite levels; it is fixed by
Eq.~\eqref{eq:triple-exact-wigner}, but is not a classical invariant.  For
the phase branch $n=0$, regular intrabasin corrections give
\begin{equation}
 \Lambda_{0l}^{(i)}=+\ii l\omega
 -\frac{B_il^2}{2\mu_iS^2}+O(S^{-3})
 +O(e^{-S^2\Delta\Phi_i}).
 \label{eq:triple-local-phase}
\end{equation}
The Hamiltonian contributes exactly $+\ii l\omega$ in angular sector $l$;
the dissipative local corrections are real before accidental level
collisions.  Local radial widths remain $O(1)$ in $r$, whereas ring radii
and their separation are $O(S)$.  For the parameters used here,
\begin{align}
 B_1&=6.25,&\sigma_1&=2.5,&
 \frac{B_1}{2\mu_1}&=6.25,
 \nonumber\\
 B_3&=42.75,&\sigma_3&=\sqrt{14.25},&
 \frac{B_3}{2\mu_3}&=14.25.
 \label{eq:triple-local-numbers}
\end{align}
Equivalently, the peaks occur near $n_i=\mu_iS^2/2$, have $O(S)$
occupation widths, and become sharply separated in $n/S^2$.

\paragraph{Stationarity, switching, and nonlocal modes.---}
At any finite Fock cutoff, trace preservation gives an exact left zero mode,
and the irreducible four-step population chain, constructed from the charge-block matrix elements in Appendix~\ref{app:numerics}, has one stationary right mode
$\Lambda_0=0$.  In the large-$S$ bistable regime its Wigner function has the
form
\begin{equation}
 W_{\rm ss}\simeq\pi_1W_{00}^{(1),R}+\pi_3W_{00}^{(3),R},
 \qquad \pi_1+\pi_3=1,
 \label{eq:triple-stationary-mixture}
\end{equation}
up to intrabasin corrections.  The other combination is the switching mode,
\begin{equation}
 W_{\rm sw}^R\simeq W_{00}^{(1),R}-W_{00}^{(3),R},
 \qquad
 \Lambda_{\rm sw}\simeq-(\Gamma_{1\to3}+\Gamma_{3\to1}),
 \label{eq:triple-switching-combination}
\end{equation}
while its left eigenfunction is approximately constant in each basin and
changes across $s_2$.  Therefore the right switching mode is not localized
at the separatrix: it has two signed lobes on the stable rings.  Only the
transition current and the derivative of the left committor form a narrow
separatrix layer.

The diffusion truncation of the scaled Wigner generator gives the
zero-current eikonal equation
\begin{align}
 \Phi'(s)&=-\frac{2A(s)}{B(s)},\\
 \Delta\Phi_i&=\int_{s_i}^{s_2}-\frac{2A(s)}{B(s)}\,\dd s,
 &\Gamma_i&\propto e^{-S^2\Delta\Phi_i}.
 \label{eq:triple-quasipotential}
\end{align}
For $(\mu_1,\mu_2,\mu_3)=(0.5,1,1.5)$, $e_1=3$,
$e_2=2.75$, and $e_3=0.75$, so that, after $u=s^2$,
\begin{align}
 Q(u)&=(u-0.5)(u-1)(u-1.5),\nonumber\\
 D(u)&=0.75+5.5u+9u^2+4u^3,\nonumber\\
 \Delta\Phi_1&=\int_{0.5}^{1}
 \frac{Q(u)}{D(u)}\,\dd u
 =1.455317761\times10^{-3},
 \nonumber\\
 \Delta\Phi_3&=\int_{1.5}^{1}
 \frac{Q(u)}{D(u)}\,\dd u
 =5.267705561\times10^{-4}.
 \label{eq:triple-barriers-diffusion}
\end{align}
Thus this reduction predicts
$-\Re\Lambda_{\rm sw}\asymp
\exp(-S^2\Delta\Phi_3)$, because the smaller outer-to-separatrix barrier
dominates the sum of rates.

There is also a sharper formal WKB statement available directly from the
exact $l=0$ population process defined by the loss and gain transitions in Appendix~\ref{app:numerics}.  With $x=n/S^2=s^2/2$, its leading
transition rates $S^2w_\nu(x)$ have jumps
\begin{align}
 (\nu,w_\nu)={}&(1,2e_3x),\quad(-2,2e_2x^2),
 \nonumber\\
 &(3,\tfrac{8e_1}{3}x^3),\quad(-4,4x^4).
 \label{eq:triple-jump-rates}
\end{align}
The activation momentum is the nonzero root of
\begin{align}
 \mathcal H(x,p)=\sum_\nu w_\nu(x)(e^{\nu p}-1)=0,
 \nonumber\\
 \Delta\Psi_i=\int_{x_i}^{x_2}p_{\rm a}(x)\,\dd x.
 \label{eq:triple-full-wkb}
\end{align}
It gives
\begin{align}
 \Delta\Psi_1&=1.455024281\times10^{-3},\\
 \Delta\Psi_3&=5.267365177\times10^{-4}.
 \label{eq:triple-full-wkb-values}
\end{align}
The near agreement with Eq.~\eqref{eq:triple-barriers-diffusion} shows that
the activation momentum is small for these parameters.  Nevertheless,
Eqs.~\eqref{eq:triple-quasipotential} and \eqref{eq:triple-full-wkb} are
eikonal estimates, not exact eigenvalues or prefactors.  A rare-event ansatz
has derivatives of order $S^2$, so every nominally higher Kramers--Moyal term
in the exact Wigner equation can contribute to its Hamilton--Jacobi equation.
The diffusion barrier must therefore not be advertised as an exact
consequence of the classical drift alone.

Linearization at the unstable ring gives
\begin{align}
 A(s_2+y)&=\kappa_2y+O(y^2),\\
 \kappa_2=A'(s_2)&=
 2\mu_2(\mu_2-\mu_1)(\mu_3-\mu_2)=0.5.
 \label{eq:triple-unstable-gap}
\end{align}
This is an inverted, not a confining, Ornstein--Uhlenbeck problem.  It has no
normalizable right Hermite ladder on the full line.  A numerical mode whose
weight lies near $s_2$ should instead be interpreted as a separatrix boundary
layer or a global resonance, and no universal formula $-n\kappa_2$ follows.
Likewise, modes extending through both basins solve a global matching problem.
Only fixed radial excitations with width
$\sigma_i\sqrt{2n+1}\ll S|s_2-s_i|$ belong to a stable-ring ladder.  Levels
with $n$ growing toward $S^2$, and accidental coincidences such as
$3\Delta_1=\Delta_3$, require global or degenerate matching.
More generally, projecting a nearly degenerate pair of inner- and outer-ring
modes onto their two local boundary layers produces a $2\times2$ matching
matrix whose off-diagonal entries are exponentially small.  Away from a
degeneracy the eigenfunctions remain localized; at a degeneracy they can form
global hybridized combinations.  The $n=l=0$ instance of this construction
is special because trace preservation pins one combination at zero and leaves
the other at the switching rate in
Eq.~\eqref{eq:triple-switching-combination}.

\paragraph{Identification of numerical modes.---}
Figure~\ref{fig:triple-main} summarizes the stationary and switching profiles, the population-sector decay rates, and the scaling of the switching mode discussed below.
For these parameters, the two local $l=0$, $n=1$ predictions are $-0.5$ and
$-1.5$; more generally the two leading ladders are
$-0.5n+\ii l$ and $-1.5n+\ii l$.  At $S=70$ the numerical values are
\begin{align}
 \Lambda_{\rm sw}&=-0.0052295,
 &\Lambda_{\rm in}&=-0.502326,
 \nonumber\\
 \Lambda_{\rm out}&=-1.497744.
 \label{eq:triple-numerical-modes}
\end{align}
The first local assignment is unambiguous by rate and localization.  The
second must be made from its outer-ring weight, not from its eigenvalue alone,
because it is asymptotically degenerate with the inner $n=3$ level.  Other
values in the numerical spectrum should not be forced onto either ladder
unless their eigenfunctions satisfy the localization and nodal tests below.
The slowest numerical $l=-1$ mode is
$-0.00133145-\ii$ and has $0.999828$ of its discrete squared weight inside
the separatrix.  It is therefore the inner $n=0$, $l=-1$ phase mode, whose
local prediction is $-6.25/S^2-\ii=-0.00127551-\ii$ at $S=70$.
No separate outer $l=-1$ branch is claimed without an outer-localized profile.
All three eigenvalues in Eq.~\eqref{eq:triple-numerical-modes} use $N=4985$;
their stored relative residuals are below $6.2\times10^{-12}$, and the
population-block trace error is $2.2\times10^{-11}$.  A least-squares fit of
all 21 integer values $60\leq S\leq80$ gives the switching exponent
$(5.417\pm0.021)\times10^{-4}$ with $R^2=0.99971$.  This fitted slope is about
$2.8\%$ above both WKB barriers, whereas the endpoint slope over $70\leq
S\leq80$ is $5.278\times10^{-4}$ and is already much closer to them.  The
window dependence is larger than the formal regression error, so the available
sizes support the exponential scale but do not distinguish the two barriers.

For a numerical Wigner eigenfunction, define the rotationally invariant
radial weight and its coordinate-invariant normalization by
\begin{equation}
 P_k(r)=\int_0^{2\pi}|W_k(r,\phi)|^2\,\dd\phi,
 \qquad
 p_k(r)=\frac{rP_k(r)}{\int_0^\infty rP_k(r)\,\dd r}.
 \label{eq:triple-radial-weight}
\end{equation}
The factor $r$ is required by the planar phase-space measure.  Choose
nonoverlapping windows
$\mathcal I_i(L)=\{|r-Ss_i|<L\sigma_i\}$ and
$\mathcal I_2(L)=\{|r-Ss_2|<L\sigma_2\}$, where $1\ll L\ll S$ and, for a
separatrix diagnostic only,
$\sigma_2^2=B(s_2)/(2\kappa_2)$.  Then set
\begin{equation}
 \mathcal W_{k,j}=\int_{\mathcal I_j(L)}p_k(r)\,\dd r,
 \qquad j=1,2,3.
 \label{eq:triple-window-weights}
\end{equation}
As $S$ grows, an inner (outer) local mode has
$\mathcal W_{k,1}\to1$ ($\mathcal W_{k,3}\to1$), an $n$th local radial
mode has $n$ radial sign changes in the signed Wigner profile within that
window, and a separatrix layer has $\mathcal W_{k,2}\to1$.  A switching or
global mode has order-one weight in both stable windows.  Since $P_k$ removes
sign and phase, it cannot by itself distinguish the stationary mixture from
the switching difference; for that one must inspect the signed $l=0$ profile
or the phase of an $l\ne0$ profile.  These tests distinguish local ladders,
higher radial excitations, separatrix layers, and basin-connecting modes
without assuming that the local expansion is spectrally complete.

To summarize the status of the claims: the Bopp generator, the exact zero
mode at finite cutoff, the angular shift $+\ii l\omega$, and the population
transition rates are exact algebraic results; the charge-sector statements are implemented by the formulas in Appendix~\ref{app:numerics}.  The Hermite functions and
$-n\Delta_i+\ii l\omega$ ladders are controlled large-$S$ results for fixed
$(n,l)$ localized near a stable ring.  The quasipotentials and exponential
switching laws are WKB estimates (with the diffusion and full-jump versions
explicitly distinguished).  The three numbers in
Eq.~\eqref{eq:triple-numerical-modes}, their localization, and the observed
switching-rate fit are numerical observations.

\section{Discussion and Outlook}

We have constructed a self-contained map from a polynomial classical drift $h(\alpha,\alpha^*)$ to explicit Hamiltonians and collapse operators.  The prescription evaluates the adjoint action exactly, applies mean-field factorization, and discards terms suppressed when $|\alpha|\sim S\to\infty$.  It thereby reproduces the prescribed classical flow while retaining GKSL form, without the recursive cancellation of lower-degree terms required by exact finite-$S$ drift matching.

Liouvillian spectroscopy tests this correspondence beyond the first-moment equation.  The fixed-point spectrum reproduces the classical stability exponents; below the Hopf threshold the origin is gapped; above it, the radial decay rate approaches the classical Floquet exponent while the phase decay vanishes.  Exactly at threshold, the minimal completion used here is nonprimitive because two-photon loss leaves even- and odd-parity dark populations, so we do not claim a universal critical-gap law for this completion.  In the bistable-ring model, two local modes recover the two classical radial relaxation rates while inter-ring switching is suppressed.  The fixed-point rapidity lattice and slow limit-cycle branches are established general phenomena \cite{DuttaZhangHaque2025}; here they serve as independent benchmarks for the explicit GKSL realization.

The scope of the construction is deliberately asymptotic.  A classical drift specifies neither an operator ordering nor the connected moments discarded by mean-field factorization.  Once the leading $O(S)$ drift is fixed, lower-degree operator terms cannot be interpreted as unique errors of the classical-to-quantum map, and canceling them in one ordering does not control other finite-$S$ effects.  Moreover, the algebraic suppression of ordering contractions does not by itself justify moment closure for arbitrary states; that step requires semiclassical localization.  Accordingly, the numerical examples test the classical limit rather than claim a uniquely determined finite-$S$ noise strength, stationary width, or subleading spectrum.

Several extensions follow naturally from this viewpoint.  The restriction to polynomial $h$ could first be relaxed to more general smooth or analytic vector fields.  On a compact region of phase space, such a field may be approximated by a sequence of polynomials and quantized term by term; an important question is then whether the resulting GKSL generators converge in a useful sense and whether the classical approximation can be made uniform on the relevant invariant sets.  Rational, piecewise-smooth, or explicitly time-dependent flows would require additional constructions, but would substantially enlarge the class of dissipative phenomena accessible to this program.

A second direction is to replace the planar phase space by other manifolds.  Of particular interest is dynamics on the sphere $\mathbb{S}^2$, whose natural quantum counterpart is a finite-dimensional spin system.  Spin coherent states and the large-spin limit $j\to\infty$ provide analogues of the bosonic amplitude and the limit $S\to\infty$, while the curvature and topology of the sphere introduce constraints absent from the plane.  A constructive correspondence on $\mathbb{S}^2$ would connect classical flows on the Bloch sphere to driven and dissipative collective-spin models and could be useful for reservoir engineering in experimentally relevant finite-dimensional systems.

The same principle should also extend to higher-dimensional classical systems through multiple bosonic modes, coupled spins, or hybrid phase spaces.  Higher dimensions support dynamical structures unavailable to autonomous planar flows, including quasiperiodicity, strange attractors, and deterministic chaos.  Constructing open quantum systems with prescribed chaotic classical limits would offer a bottom-up route to quantum dissipative chaos.  In that setting, classical Lyapunov exponents, invariant measures, and decay resonances could be compared with quantum trajectories, phase-space distributions, and the low-lying Liouvillian spectrum, clarifying which signatures of chaos survive at finite effective Planck constant and how quantum noise modifies them.

Finally, the nonuniqueness of the quantum completion can be treated as a resource rather than merely a limitation.  Once the leading classical drift is fixed, the remaining freedom in jump operators may be used to impose symmetries, locality, experimentally available couplings, desired fluctuation statistics, or reduced channel complexity.  This suggests an inverse-design problem: among the family of GKSL generators with the same classical limit, select representatives optimized for physical implementation or for a chosen finite-$S$ quantum effect.  Together, these directions may turn classical dynamical systems into a broader design language for quantum open-system dynamics.

\section*{Acknowledgments}
Tingfei Li acknowledges support from the National Natural Science Foundation of China (Grant No.~12605129). 
\appendix

\section{Normal-ordered algebra and leading mixed terms}
\label{app:algebra}

For normal-ordered monomials, repeated use of $aa^\dagger=a^\dagger a+1$ gives
\begin{equation}
 a^q(a^\dagger)^r=
 \sum_{\ell=0}^{\min(q,r)}
 \binom{q}{\ell}\frac{r!}{(r-\ell)!}
 (a^\dagger)^{r-\ell}a^{q-\ell}.
 \label{eq:normal-product}
\end{equation}
Every contraction lowers the total degree by two. The code in \texttt{src/operator\_poly.py} implements Eq.~\eqref{eq:normal-product} exactly.

For
\begin{equation}
 H=z(a^\dagger)^pa^q+z^*(a^\dagger)^qa^p,
\end{equation}
we use $[(a^\dagger)^p,a]=-p(a^\dagger)^{p-1}$ to obtain
\begin{equation}
 \ii[H,a]=-
 \ii pz(a^\dagger)^{p-1}a^q-
 \ii qz^*(a^\dagger)^{q-1}a^p.
 \label{eq:Haction-app}
\end{equation}
This identity is exact because the annihilation powers commute with $a$. For later use, define the bilinear adjoint action
\begin{equation}
	\mathcal{D}^\dagger_{L_1,L_2}(X)
	=\frac12\left(L_1^\dagger[X,L_2]+[L_1^\dagger,X]L_2\right),
	\label{eq:bilinear-dissipator}
\end{equation}
so that $\D_L^\dagger(X)=\mathcal{D}^\dagger_{L,L}(X)$. With the bilinear notation $\mathcal D^\dagger_{L_1,L_2}$ introduced above, the standard adjoint dissipator can be written
\begin{equation}
 \D_L^\dagger(O)=\mathcal D^\dagger_{L,L}(O)
 =\frac12\left(L^\dagger[O,L]+[L^\dagger,O]L\right).
 \label{eq:Dcomm-form}
\end{equation}
For $L=(a^\dagger)^pa^q$, retaining the no-contraction term after the elementary commutators gives
\begin{align}
 \D_L^\dagger(a)={}&
 \frac{p-q}{2}(a^\dagger)^{p+q-1}a^{p+q}\nonumber\\
 &+\text{terms lower by at least two degrees}.
 \label{eq:Dmono-leading}
\end{align}

Let
\begin{equation}
 A=(a^\dagger)^pa^q,
 \qquad
 B=\xi(a^\dagger)^ra^s.
\end{equation}
The mixed part of $\D_{A+B}^\dagger(a)$ is
\begin{align}
 \mathcal D_{A,B}^\dagger(a)+\mathcal D_{B,A}^\dagger(a)
 ={}&\frac12\left(A^\dagger[a,B]+[A^\dagger,a]B\right)
 \nonumber\\
 &+\frac12\left(B^\dagger[a,A]+[B^\dagger,a]A\right).
 \label{eq:mixed-def}
\end{align}
Using the no-contraction part of Eq.~\eqref{eq:normal-product},
\begin{align}
 \mathcal D_{A,B}^\dagger(a)+\mathcal D_{B,A}^\dagger(a)
 \simeq{}&\frac{\xi(r-q)}{2}
 (a^\dagger)^{r+q-1}a^{p+s}
 \nonumber\\
 &+\frac{\xi^*(p-s)}{2}
 (a^\dagger)^{p+s-1}a^{r+q}.
 \label{eq:mixed-leading}
\end{align}
The symbol $\simeq$ in this appendix means equality of the highest total normal-ordered degree only. The full code never drops the lower-degree contractions.

\section{Detailed derivation of the even block}
\label{app:even}

Set $n=2m$ and
\begin{equation}
 A=a^{m+1},
 \qquad
 B=\xi(a^\dagger)^{m-k-1}a^{k+1},
 \qquad
 G=(a^\dagger)^{m+1}.
 \label{eq:even-ABG}
\end{equation}
For each $k$, the algebraic block uses $J_1=\sqrt{2w}(A+B)$ and one copy of $\sqrt{2w}G$, where $w=S^{1-2m}$. Summing the $m$ identical gain dissipators gives the single consolidated operator $J^{(g)}_{2m}$ in Eq.~\eqref{eq:even-Jg}. The factor $2w$ cancels the factor $1/2$ in Eq.~\eqref{eq:mixed-leading}.

For $A$, Eq.~\eqref{eq:Dmono-leading} gives
\begin{equation}
 2w\D_A^\dagger(a)\simeq-w(m+1)(a^\dagger)^ma^{m+1}.
\end{equation}
For $G$,
\begin{equation}
 2w\D_G^\dagger(a)\simeq+w(m+1)(a^\dagger)^ma^{m+1}.
\end{equation}
Thus their leading self-actions cancel. The lower-degree self-action of $B$ is discarded.

In Eq.~\eqref{eq:mixed-leading}, choose $p=0$, $q=m+1$, $r=m-k-1$, and $s=k+1$. The two cross terms are
\begin{align}
 2w\bigl[\mathcal D_{A,B}^\dagger(a)+\mathcal D_{B,A}^\dagger(a)\bigr]
 \simeq{}&-w(k+2)\xi(a^\dagger)^{2m-k-1}a^{k+1}\nonumber\\
 &-w(k+1)\xi^*(a^\dagger)^ka^{2m-k}.
 \label{eq:even-cross}
\end{align}
The Hamiltonian in Eq.~\eqref{eq:even-H} gives
\begin{align}
 \ii[H_{2m,k},a]={}&-
 \frac{\ii(k+1)z}{S^{2m-1}}(a^\dagger)^ka^{2m-k}
 \nonumber\\
 &-\frac{\ii(2m-k)z^*}{S^{2m-1}}
 (a^\dagger)^{2m-k-1}a^{k+1}.
 \label{eq:even-Haction}
\end{align}
Adding Eqs.~\eqref{eq:even-cross} and \eqref{eq:even-Haction} yields Eq.~\eqref{eq:even-match}.

To solve it, conjugate the second equation:
\begin{equation}
 -(k+2)\xi^*+\ii(2m-k)z=\nu^*.
\end{equation}
Multiply the first equation by $k+2$, the conjugated equation by $k+1$, and subtract. The $\xi^*$ terms cancel and one obtains
\begin{equation}
 -2\ii(k+1)(m+1)z=(k+2)\mu-(k+1)\nu^*,
\end{equation}
which gives Eq.~\eqref{eq:even-z}. Substitution into the conjugate of the first matching equation gives Eq.~\eqref{eq:even-xi}. The determinant of the matching system is $2(k+1)(m+1)$ and is nonzero over the full allowed index range.

All omitted terms in the exact normal ordering have lower degree than the target and are discarded in the large-$S$ mean-field limit.

\section{Detailed derivation of the odd block and central completion}
\label{app:odd}

Set $n=2m+1$ and
\begin{equation}
 A=a^{m+1},
 \qquad
 B=\xi(a^\dagger)^{m-k}a^{k+1},
 \qquad
 w=S^{-2m}.
 \label{eq:odd-AB}
\end{equation}
Both $A$ and $B$ have total degree $m+1$. For $J=\sqrt{2w}(A+B)$, Eq.~\eqref{eq:mixed-leading} gives
\begin{align}
 \text{cross}(\D_J^\dagger a)\simeq{}&-w(k+1)\xi^*
 (a^\dagger)^ka^{2m+1-k}
 \nonumber\\
 &-w(k+1)\xi(a^\dagger)^{2m-k}a^{k+1}.
 \label{eq:odd-cross}
\end{align}
The Hamiltonian contribution is
\begin{align}
 \ii[H_{2m+1,k},a]={}&-
 \frac{\ii(k+1)z}{S^{2m}}(a^\dagger)^ka^{2m+1-k}
 \nonumber\\
 &-\frac{\ii(2m+1-k)z^*}{S^{2m}}
 (a^\dagger)^{2m-k}a^{k+1}.
 \label{eq:odd-Haction}
\end{align}
Equating coefficients gives Eq.~\eqref{eq:odd-match}. Conjugating the second equation and subtracting it from the first eliminates $\xi^*$:
\begin{equation}
 -2\ii(m+1)z=\mu-\nu^*,
\end{equation}
which gives Eq.~\eqref{eq:odd-z}; back-substitution gives Eq.~\eqref{eq:odd-xi}.

The self-action of $A$ is
\begin{equation}
 2w\D_A^\dagger(a)\simeq-w(m+1)(a^\dagger)^ma^{m+1}.
\end{equation}
For $B$, its creation and annihilation powers are $p=m-k$ and $q=k+1$. Therefore
\begin{equation}
 2w\D_B^\dagger(a)\simeq
 w|\xi|^2(m-2k-1)(a^\dagger)^ma^{m+1}.
\end{equation}
Their sum is $-w\beta_k(a^\dagger)^ma^{m+1}$, with $\beta_k$ in Eq.~\eqref{eq:epsilon}. Summing over $k<m$, the missing central coefficient is
\begin{equation}
 \eps=C_m+\sum_{k=0}^{m-1}\beta_k,
\end{equation}
as defined in Eq.~\eqref{eq:epsilon}.

For $N_{m+1}=(a^\dagger)^{m+1}a^{m+1}$,
\begin{equation}
 [N_{m+1},a]=-(m+1)(a^\dagger)^ma^{m+1}.
\end{equation}
Thus Eq.~\eqref{eq:central-H} contributes $\ii w\Im\eps$ to the central coefficient. A gain jump $(a^\dagger)^{m+1}$ contributes $+w\Re\eps$, while a loss jump $a^{m+1}$ contributes the same signed value when $\Re\eps<0$. Eqs.~\eqref{eq:central-Jplus} and \eqref{eq:central-Jminus} therefore add $w\eps$, and the total central coefficient is $-w\sum_k\beta_k+w\eps=wC_m$.

Because all monomials in the odd composite jump have equal total degree, every normal-ordering contraction lowers the degree. These lower-degree terms are discarded in the large-$S$ mean-field limit.

\section{Charge blocks used in the numerical tests}
\label{app:numerics}

The vectorization convention is column stacking, consistent with standard Liouville-space constructions \cite{ProsenSeligman2010,McDonaldClerk2023},
\begin{equation}
 \operatorname{vec}(A\rho B)=(B^{\mathsf T}\otimes A)\operatorname{vec}(\rho).
\end{equation}
For phase-covariant monomial jumps, use the angular index $l=-q$ and the basis $|n-l\rangle\langle n|$. The Hamiltonian contributes $+\ii\omega l$. A loss jump $\sqrt\gamma a^p$ maps
\begin{align}
 |m\rangle\langle n|\mapsto{}&
 \gamma\sqrt{m^{\underline p}n^{\underline p}}
 |m-p\rangle\langle n-p|
 \nonumber\\
 &-\frac{\gamma}{2}
 \left(m^{\underline p}+n^{\underline p}\right)|m\rangle\langle n|,
 \label{eq:loss-charge}
\end{align}
where $n^{\underline p}=n(n-1)\cdots(n-p+1)$. A gain jump replaces falling factorials by $(n+1)^{\overline p}$ and shifts both indices upward. These formulas construct an exact $(N-|l|)\times(N-|l|)$ block without forming the $N^2$ Liouville matrix.

\bibliographystyle{apsrev4-2}
\bibliography{references}

\end{document}